\documentclass{ws-mpla}
\usepackage[super]{cite}
\usepackage{graphicx}
\usepackage{amssymb}
\usepackage{amsmath}
\usepackage{amsmath,multicol}
\usepackage{tabularx,tabulary}
\usepackage{bm}
\usepackage{comment}
\usepackage{caption}
\usepackage{subcaption}
\usepackage{xcolor}
\usepackage{tcolorbox} %
\tcbset{colframe=blue, boxrule=0.5pt, arc=0mm, outer arc=0mm, boxsep=5pt, colback=white}

\def\a {\epsilon}
\def\g {\gamma}

\def\M  {{\cal M}}

\def\P  {{\cal P}\!}
\def\ap#1#2#3   {{\rm Ann. Phys. (NY)}       #1 (#3) #2}
\def\apj#1#2#3  {{\rm Astrophys. J.}         #1 (#3) #2}
\def\apjl#1#2#3 {{\rm Astrophys. J. Lett.}   #1 (#3) #2}
\def\app#1#2#3  {{\rm Acta. Phys. Pol.}      #1 (#3) #2}
\def\chp#1#2#3  {{Chin.\ Phys. }             #1 (#3) #2}
\def\cpc#1#2#3  {{\rm Computer Phys. Comm.}  #1 (#3) #2}
\def\dum#1#2#3  {{~}                         #1 (#3) #2}
\def\epjc#1#2#3 {{\rm Eur. Phys. J. C}       #1 (#3) #2}
\def\err#1#2#3  {{\it Erratum}               #1 (#3) #2}
\def\ib#1#2#3   {{\it ibid.}                 #1 (#3) #2}
\def\jcp#1#2#3  {{\rm J. Comp. Phys.}        #1 (#3) #2}
\def\jmp#1#2#3  {{\rm J. Math. Phys.}        #1 (#3) #2}
\def\jhep#1#2#3 {{\rm JHEP}                  #1 (#3) #2}
\def\ijmp#1#2#3 {{\rm Int. J. Mod. Phys.}    #1 (#3) #2}
\def\jpg#1#2#3  {{\rm J. Phys. G.}           #1 (#3) #2}
\def\mpl#1#2#3  {{\rm Mod. Phys. Lett.}      #1 (#3) #2}
\def\mpla#1#2#3 {{\rm Mod. Phys. Lett. A}    #1 (#3) #2}
\def\nat#1#2#3  {{\rm Nature (London)}       #1 (#3) #2}
\def\ncim#1#2#3 {{\rm Nuovo Cimento}         #1 (#3) #2}
\def\nca#1#2#3  {{\rm Nuovo Cimento A}       #1 (#3) #2}
\def\ncb#1#2#3  {{\rm Nuovo Cimento B}       #1 (#3) #2}
\def\nim#1#2#3  {{\rm Nucl. Instr. Meth.}    #1 (#3) #2}
\def\njp#1#2#3  {{New J. Phys. }             #1 (#3) #2}
\def\np#1#2#3   {{\rm Nucl. Phys.}           #1 (#3) #2}
\def\npb#1#2#3  {{\rm Nucl. Phys. B}         #1 (#3) #2}
\def\pan#1#2#3  {{\rm Phys. At. Nuclei}      #1 (#3) #2}
\def\pl#1#2#3   {{\rm Phys. Lett.}           #1 (#3) #2}
\def\plb#1#2#3  {{\rm Phys. Lett. B}         #1 (#3) #2}
\def\prep#1#2#3 {{\rm Phys. Rep.}            #1 (#3) #2}
\def\prev#1#2#3 {{\rm Phys. Rev.}            #1 (#3) #2}
\def\prc#1#2#3  {{\rm Phys. Rev. C}          #1 (#3) #2}
\def\prd#1#2#3  {{\rm Phys. Rev. D}          #1 (#3) #2}
\def\prev#1#2#3 {{\rm Phys. Rev.}            #1 (#3) #2}
\def\prl#1#2#3  {{\rm Phys. Rev. Lett.}      #1 (#3) #2}
\def\prs#1#2#3  {{\rm Proc. Roy. Soc.}       #1 (#3) #2}
\def\ptp#1#2#3  {{\rm Prog. Theor. Phys.}    #1 (#3) #2}
\def\ptep#1#2#3 {{\rm Prog. Theor. Exp. Phys.}  #1 (#3) #2}
\def\ps#1#2#3   {{\rm Physica Scripta}       #1 (#3) #2}
\def\rmp#1#2#3  {{\rm Rev. Mod. Phys.}       #1 (#3) #2}
\def\rpp#1#2#3  {{\rm Rep. Prog. Phys.}      #1 (#3) #2}
\def\sa#1#2#3   {{Sci.\ Acta }               #1 (#3) #2}
\def\sci#1#2#3  {{Science }                  #1 (#3) #2}
\def\sjnp#1#2#3 {{\rm Sov. J. Nucl. Phys.}   #1 (#3) #2}
\def\spj#1#2#3  {{\rm Sov. Phys. JETP}       #1 (#3) #2}
\def\spu#1#2#3  {{\rm Sov. Phys.-Usp.}       #1 (#3) #2}
\def\yaf#1#2#3  {{\rm Yad. Fiz.}             #1 (#3) #2}
\def\zp#1#2#3   {{\rm Zeit. Phys.}           #1 (#3) #2}
\def\zpa#1#2#3  {{\rm Zeit. Phys. A}         #1 (#3) #2}
\def\zpc#1#2#3  {{\rm Zeit. Phys. C}         #1 (#3) #2}

\begin{document}

\markboth{A. G. Bagdatova, S. P. Baranov}{Polarization and kinematic properties 
of the splitting functions $q\to W^\pm +q'$ and $q\to Z^0 +q$}


\title{Polarization and kinematic properties of the splitting functions \\
$q\to W^\pm +q'$ and $q\to Z^0 +q$}
\author{Alsu G. Bagdatova$^{\ast}$, ~{\rm Sergey P. Baranov}$^{\dagger}$}
\address{P.N. Lebedev Institute of Physics,
              53 Lenin Avenue, 119991 Moscow, Russia\\
$^{\ast}$~bagdatovaag@lebedev.ru\\
$^{\dagger}$~baranovsp@lebedev.ru}

\maketitle


\begin{abstract}
We consider the processes $q\to W{+}\,q'$ and $q\to Z{+}\,q$ and derive the 
respective splitting functions as functions of two kinematic variables: the 
longitudinal momentum fraction $z$ and transverse momentum $p_T$ of the produced 
bosons with respect to the parent quark. We take into account kinematic (phase 
space) restrictions connected with nonzero masses of the gauge bosons and with 
limited initial energy. We separately consider three different polarization 
states of the bosons.
\keywords{Perturbation theory; Electroweak interactions; Splitting function.} 
\end{abstract}

\ccode{PACS Nos.: 12.38.Bx, 12.15.J}

\section{Introduction}
The structure of the proton, when probed at increasing energies, reveals increasing
complexity of its composition. Not only light quarks and gluons can be found among 
the proton's constituents, but also heavy quarks c and b and, maybe, even the 
electroweak bosons $W$ and $Z$. The bosons can be present in the virtual form as sea
partons or in the real form as final state quanta. Our note focuses on the latter 
case. The emission of the final state quanta can be conveniently described in terms
of quark splitting functions $q\to W^\pm{+}\,q'$ and $q\to Z^0{+}\,q$, which we are 
going to derive in this note. 
This problem has already been addressed in the pioneering works \cite{Kane,Dawson} 
and later in Refs. \cite{Chen,Bauer,Fornal,Han}. We have, however, introduced three
innovations.

First, we consider the splitting function as a function of two (rather than one)
kinematic variables, $z$ and $p_T$, where $z$ is the $W$ longitudinal momentum 
fraction and $p_T$ the transverse momentum with respect to the momentum of the parent 
(splitting) quark. Second, we take into account kinematic restrictions, that is, phase
space limitations connected with nonzero $W$ and $Z$ masses and nonzero $p_T$. This 
may be especially important for particle event generators \cite{Sjostrand,Mas} running 
at the energies of real colliders (and not at the asymptotic energies $\sqrt{s}\to\infty$).
Third, we not only make distinction between the longitudinal and transverse 
polarizations of the $W$ or $Z$ bosons, but also between two transverse polarizations
(the polarization vector may either lie in the boson production plane or be
perpendicular to this plane). 
These issues were absent in the previous calculations known to the authors.

\section{Calculation}
\label{calc}
\begin{figure}[h]
\centering 	\includegraphics[width=0.95\textwidth]{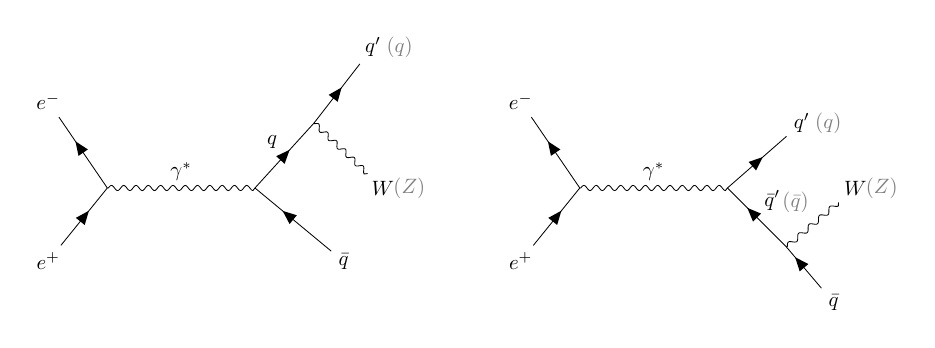}	
\caption{Feynman diagrams considered in the derivation of the $q\to Wq'$ or $q\to Zq$
splitting functions. } 
\label{fig:Feynman}%
\end{figure}

To calculate the quark to W splitting function, we start with the process 
\begin{equation}
e^+e^-\to\gamma^*\to \bar{q}+q+W \label{qqW}
\end{equation}
considered in the virtual photon rest frame.
The diagrams shown in Fig.~\ref{fig:Feynman} constitute a gauge invariant subset of all lowest order
diagrams\footnote{The other two diagrams describe the emission of the boson from the
initial leptons. They constitute another gauge invariant subset.}. 
The corresponding amplitudes read
\begin{eqnarray}
\M_1 &=& \,g_{W}\,\frac{e^2}{s}\,\a^\nu\,[\bar{v}(q_2)\,\g^\mu\,u(q_1)]\,\,[\bar{u}(p_1)\,W_\nu \,\frac{(\not{\!}\!p_W+\not{\!}\!p_1)}{(p_W{+}p_1)^2}\,\g_{\mu}\,v(p_2)], \nonumber\\
\M_2 &=& \,g_{W}\,\frac{e^2}{s}\,\a^\nu\,[\bar{v}(q_2)\,\g^\mu \,u(q_1)]\,\,[\bar{u}(p_1)\,\g_{\mu}\,\frac{(\not{\!}\!p_W+\not{\!}\!p_2)}{(p_W{+}p_2)^2}\,W_\nu\,v(p_2)] 
\end{eqnarray}

where $g_{W}=e\,V_{q\bar{q}}/\sqrt{8}\sin\theta_{W}$ is the weak coupling constant, 
$W_\nu = \g_\nu\,(1-\g_5)$ the standard W boson coupling, $q_1$ and $q_2$ denote the 
initial $e^+$ and $e^-$ momenta, $p_1$, $p_2$ and $p_W$
the final state quark, antiquark and boson momenta, $\a$ the boson polarization vector,
and $s$ the invariant mass of the whole system, $s=(q_1{+}q_2)^2=(p_1{+}p_2{+}p_W)^2$.

Throughout this paper, we use the following orientation of the coordinate axes.
In the overall center-of-mass frame, the momenta $p_1$, $p_2$ and $p_W$ lie in 
a plane which we call the production plane.
The longitudinal polarization vector $\a_z$ is oriented along the momentum of
the gauge boson. The polarization vector $\a_x$ is perpendicular to $\a_z$ and 
lies in the W production plane. The polarization vector $\a_y$ is perpendicular 
to $\a_z$ and $\a_x$.

The amplitude $\M_1$ favors the emission of bosons in the quark direction. The virtual 
quark propagator is then of the order $1/m_W^2$, while for the emission in the antiquark 
direction it is of the order $1/s$. For the same reason, the amplitude $\M_2$ favors the
emission of bosons in the antiquark direction. The interference terms are of the order 
$1/(s\,m_W^2)$ and are small in comparison with 'favored emission' probabilities of the
order $1/m_W^4$ when $s\gg m_W^2$. Then, the entire process can be factorized into the 
production of a $q\bar{q}$
pair and a $q\to q'W$ or $\bar{q}\to\bar{q}'W$ splitting.


The fully differential cross section for the process (\ref{qqW}) reads~\cite{BycKaj}
\begin{eqnarray} 
d\sigma(e^+\!e^-\!\!\to\!W\,q\,\bar{q}) = 
\frac{1}{2s}\,\frac{1}{(2\pi)^5}\,
\bigl|\M (ee\!\to \gamma^*\!\to\!W q\,\bar{q} )\,\bigr|^2
\;\frac{ds_1\,ds_2\,d\phi\,d\psi\,d\cos\theta}{32s},
\label{dsigma}
\end{eqnarray} 
where $s_1=(p_1+p_W)^2$, $s_2=(p_2+p_W)^2$, and $\phi$, $\psi$\textcolor{black}{,} $\theta$ are three 
Euler's angles describing the orientation of the coordinate system. The differential 
cross section of the 'prequel' process (before the quark splitting) is
\begin{eqnarray}
d\sigma(e^+e^-\!\!\to q\,\bar{q}) =
\frac{1}{2s}\,\frac{1}{(2\pi)^2}\,
\bigl|\M (ee\!\to\!\gamma^*\!\to\! q\,\bar{q})\,\bigr|^2\;
\frac{\lambda^{1/2}(s,p^{*2}\!,m_q^2)}{8s}\;d\Omega,
 \label{qq}
\end{eqnarray}
where 
$\lambda^{1/2}(x,y,z) = (x^2 + y^2 + z^2 - 2xy - 2xz - 2yz)^{1/2}$
and $p^*$ is the momentum of the parent (splitting) quark. To decide whether the
generated event corresponds to quark splitting case or antiquark splitting case, 
we compare the values of $s_1$ and $s_2$ and make choice in favor of the smallest of 
them. That is, we set $p^*=p_1+p_W$ if $s_1 < s_2$, and $p^*=p_2+p_W$ if $s_1 > s_2$.
As an alternative, one could consider the angular separation 
$\Delta R=(\Delta\eta^2 +\Delta\phi^2)^{1/2}$, where $\Delta\eta$ is the difference 
between the pseudorapidities of the quark (or antiquark) and the gauge boson, and 
$\Delta\phi$ is the difference between the azimuthal angles. Then, the splitting 
quark $p^*$ is the one which shows the smallest $\Delta R$. The results based on
these two 
definitions of the splitting subsystem are very close numerically.

The calculation of Feynman diagrams is straightforward and follows standard QED/QCD 
rules. All calculations were done using the algebraic manipulation system {\sc form}
\cite{FORM}.

After dividing Eq.(\ref{dsigma}) by Eq.(\ref{qq}) 
we arrive at the definition of the fully differential splitting function
\begin{eqnarray}
d\P\,(q^*\!\!\to\!Wq)\!=\!
\frac{1}{(2\pi)^3}
\frac{1}{4\,\lambda^{1/2}(s,p^{*2}\!,m_q^2)}\;
\frac{\bigl|\M (\gamma^*\!\!\to W\,q\,\bar{q})\bigr|^2}
{\bigl|\M (\gamma^*\!\!\to q\,\bar{q})\bigr|^2}\,
ds_1\,ds_2\,d\phi\,d\psi\,d\cos\theta.
\label{Dz}
\end{eqnarray}
The latter can be reduced to the conventional splitting function $\P_{q/W}(z)$ 
by introducing the light-cone variable $z=p^+_W/p^{*+}=(E_W+p_{W,||})/(E^*+|p^*|)$ 
and integrating over all other variables in Eq.(\ref{Dz}):
\begin{eqnarray}
\P_{q/W}(z) = \int\!\! \P\,(q^*\to W\,q)\;\delta(z - p^+_W/p^{*+})\;ds_1\,ds_2\;
d\phi\,d\psi\;d\cos\theta.
\end{eqnarray}
We also consider splitting function with un-integrated $p_T$ dependence: $\P\,(z,p_T;s)$.

The integration limits for the angular variables $d\phi\,d\psi\,d\cos\theta$ are trivial,
while the ones for $ds_1\,ds_2$ are not. They are determined by the requirement that 
the process be lying in the physical region of the phase space \cite{BycKaj}: 
\begin{equation}
G(s_2,\,s_1,\,m_{\bar{q}}^2,\,m_q^2,\,s,\,m_W^2) < 0, \label{gf}
\end{equation}
where the function $G$ is defined as (eq. 5.23 in \cite{BycKaj})
\begin{eqnarray}
G(x,y,z,u,v,w)= xzw + xuv + yzv + yuw -xy\,(z+u+v+w-x-y) \nonumber\\
-zu\,(x+y+v+w-z-u) -vw\,(x+y+z+u-v-w).
\end{eqnarray}

The integration was performed by means of the Monte-Carlo technique, using the routine 
{\sc vegas} \cite{VEGAS}. 

The complexity of the phase space boundary does not allow us to present our results
in the form of compact analytic expressions. We have, however, turned our calculations
into a {\sc fortran} and a {\sc C++} codes which are made public\footnote{These codes 
are available from the authors on request} and can be conveniently incorporated in 
Monte Carlo event generators.

We emphasize that the phase space limitations make the splitting function $\P\,(z,p_T;s)$ 
scale dependent. The invariant energy $s$ plays the role of probing scale. At low $s$, 
the scale dependence mostly comes from the kinematics, that is, from a requirement that
the quark energy be large enough to produce a heavy boson. At much higher energies, the 
scale dependence is dominated by radiative corrections (not considered in this note), 
that is, by multiple gluon emission from quarks. 

\section{Numerical results}
\label{show}
The behavior of all splitting functions is qualitatively very similar. To be definite, we
show the results obtained for quark to $W$ splitting, $q\to Wq'$. To come from $W$ to $Z$ 
one only has to change the boson mass (resp., $m_W$ or $m_Z$) and introduce an overall
normalizing factor $(1+C_u^2\,)/(4\cos^2\theta_W)$ for $u\to Zu$, 
               and $(1+C_d^2\,)/(4\cos^2\theta_W)$ for $d\to Zd$,
with $C_u=1-(8/3)\sin^2\theta_W$  and $C_d=1-(4/3)\sin^2\theta_W$. The light quark 
mass can be safely set to zero, $m_q=0$.

\begin{figure}
\centering 	
\includegraphics[width=0.7\textwidth]{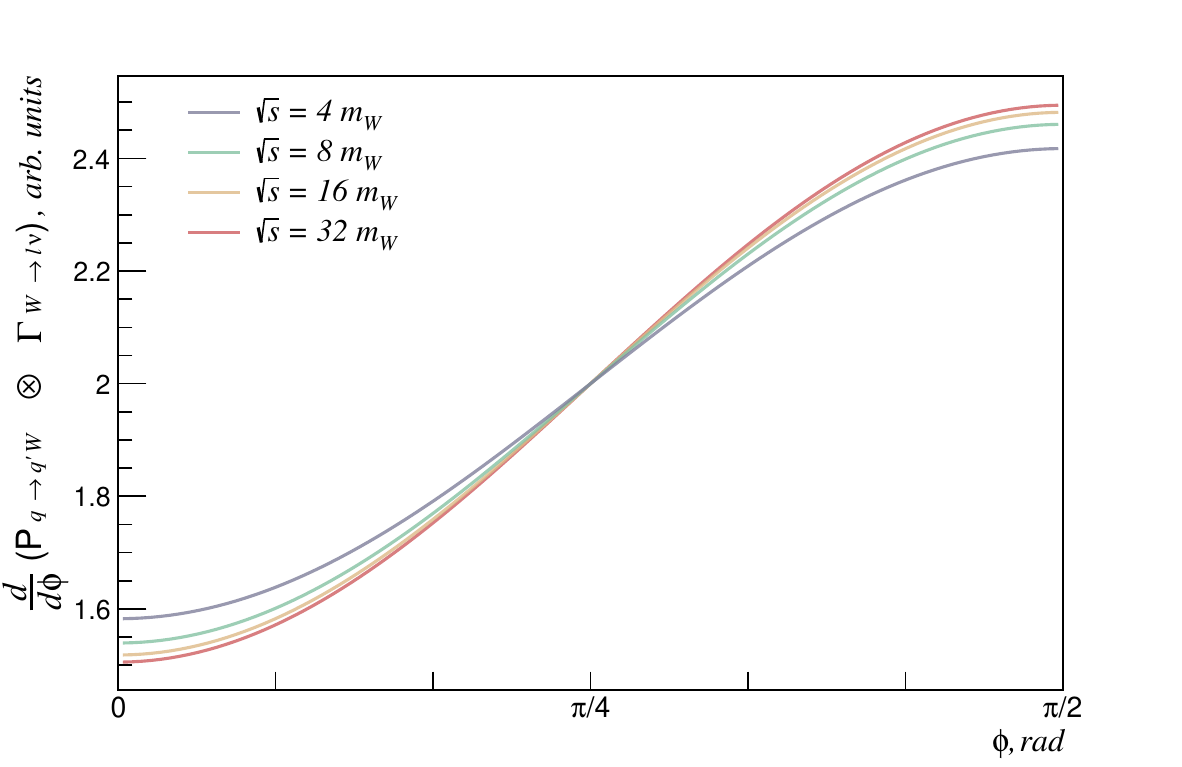}	
\caption{Azimuthal angle distributions between the W production plane and the W decay plane.
The curves are normalized to have the same integral area (or, equivalently, to have 
the same value at $\phi=\pi/4$).}
\label{fig:angle}%
\end{figure}

We draw the reader's attention to the non-equivalence of two transverse polarizations 
$\a_x$ and $\a_y$, when the $W$ polarization vector may lie in the $W$ production plane
or be perpendicular to this plane.
The non-equivalence of two polarization states would not open up to one's eye when 
using the helicity basis $\a_{\pm} = (\pm\a_x +i\a_y)/\sqrt{2}$, because the positive
and negative helicity states are produced with strictly equal probabilities 
$|\M_+|^2 = |\M_-|^2 =\left( |\M_x|^2 + |\M_y|^2\right)/2$.
However, the physical effect does exist, and it can be detected experimentally through
measuring the azimuthal asymmetry of the W decay products, as it is shown 
in Fig.~\ref{fig:angle} where we plot distributions over the angle between the 
W production plane and the W decay plane. The ordinary splitting functions would predict
uniform angular distributions as they are indifferent to the production plane.

\textcolor{black}{The shape of the angular distribution may look counter-intuitive. Given 
the fact that the gauge bosons are predominantly polarized in the production plane, one 
could expect a similar correlation for the decay plane as well. It would exactly be so 
if the quark-boson coupling was either of purely vector or of purely axial-vector type. 
However, the interference between the vector and axial-vector terms inverts the angular 
dependence and makes the production and decay planes perpendicular to each other.}

\begin{figure}
\centering
\includegraphics[width=0.49\textwidth]{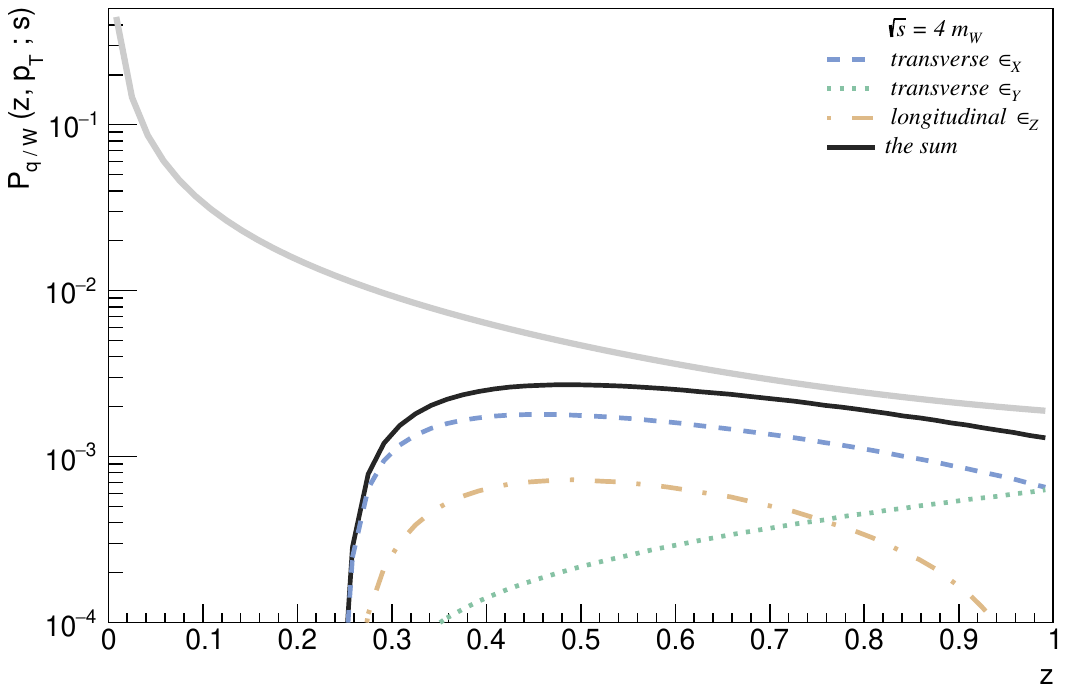}
\includegraphics[width=0.49\textwidth]{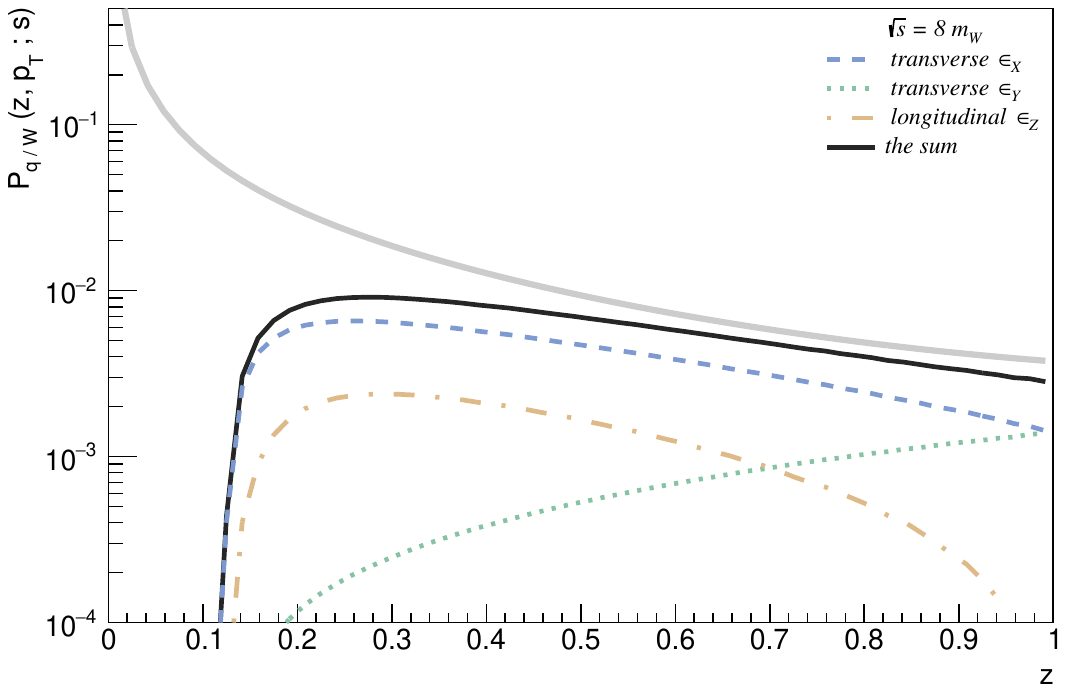}
\includegraphics[width=0.49\textwidth]{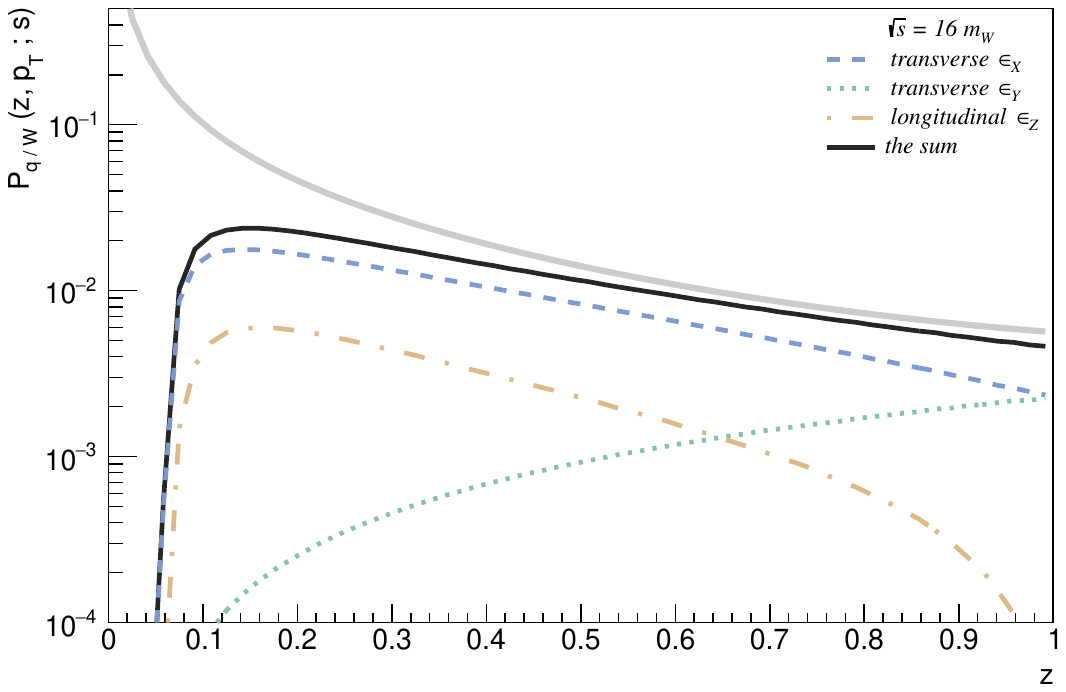}
\includegraphics[width=0.49\textwidth]{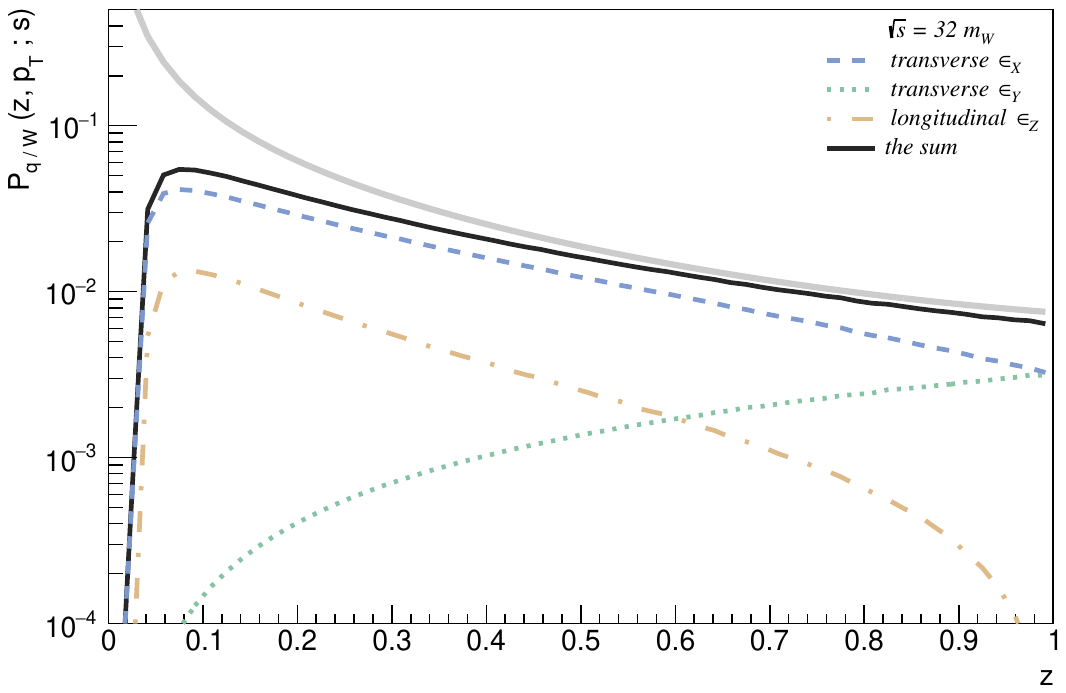}
\caption{
Net longitudinal momentum spectra obtained by integrating the splitting function 
$\P_{q/W}\,(z,p_T;s)$ over the allowed $p_T$ range for four different values of $s$.
Dashed curves, transverse polarization $\epsilon_{x}$ in the $Wq$ production plane;
dotted curves, transverse polarization $\epsilon_{y}$ perpendicular to the $Wq$ 
production plane; dash-dotted curves, longitudinal polarization $\epsilon_{z}$;
solid curves, the sum of all contributions. 
The values of $s$ from top to bottom: $\sqrt{s}~=~4\,m_W,\,8\,m_W,\,16\,m_W,\,32\,m_W$.
For comparison we also plot the function
$\P\,_{\textcolor{black}{q/W}}\,(z,s) = \displaystyle{\frac{\alpha_{\mbox{\tiny\rm eff}}}{2\pi}\,
\Bigl[\frac{1+(1-z)^2}{z}\,\ln\!\frac{s}{4m_W^2}\Bigr]}$
with $\alpha_{\mbox{\tiny\rm eff}}=\displaystyle{\frac{\alpha}{4\sin^2\!\theta_W}}$
that represents Weizs\"{a}cker-Williams approximation for W bosons \cite{Sjostrand}, 
gray solid curves.}
\label{fig:fz}
\end{figure}

\begin{figure}
\centering
\includegraphics[width=0.49\textwidth]{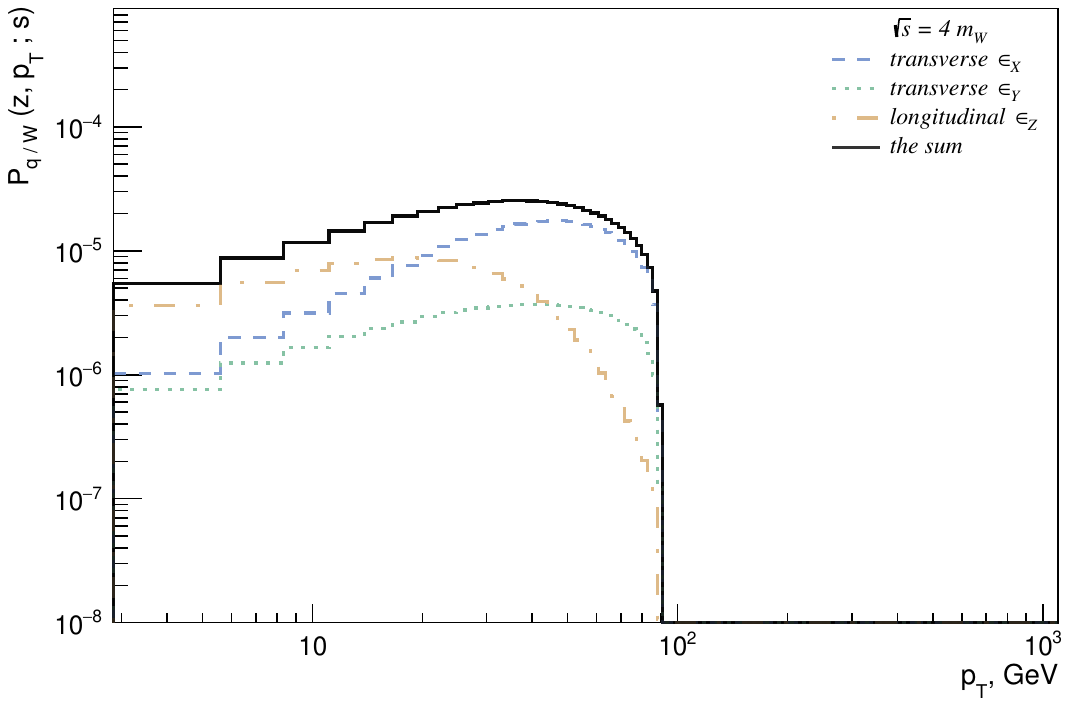}
\includegraphics[width=0.49\textwidth]{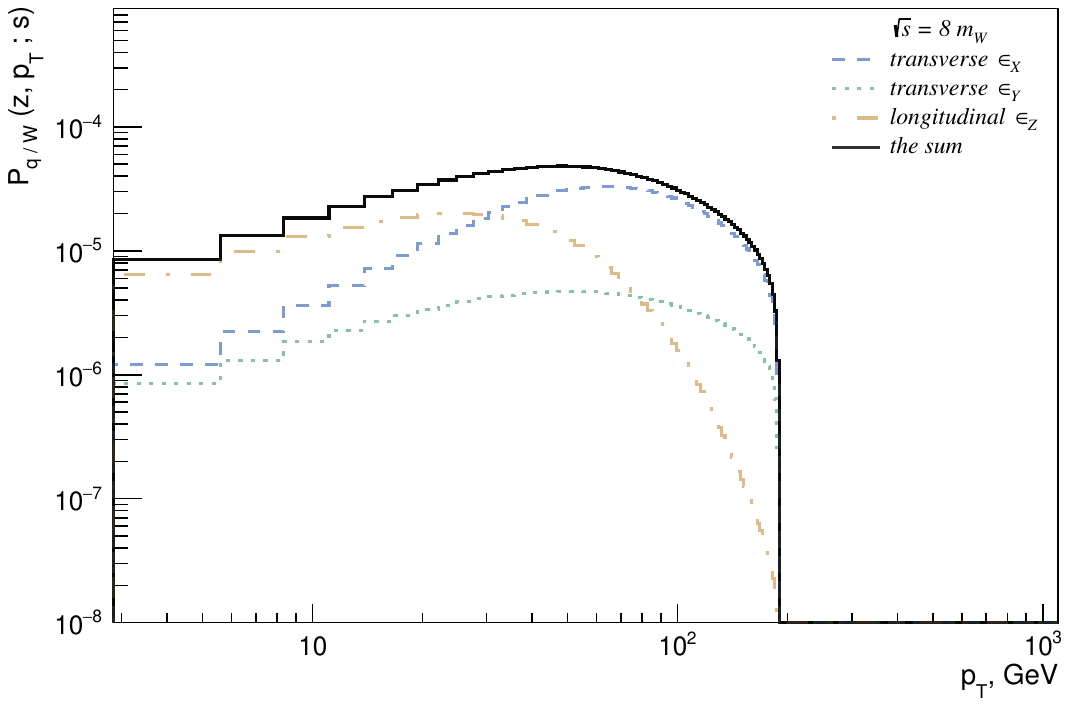}
\includegraphics[width=0.49\textwidth]{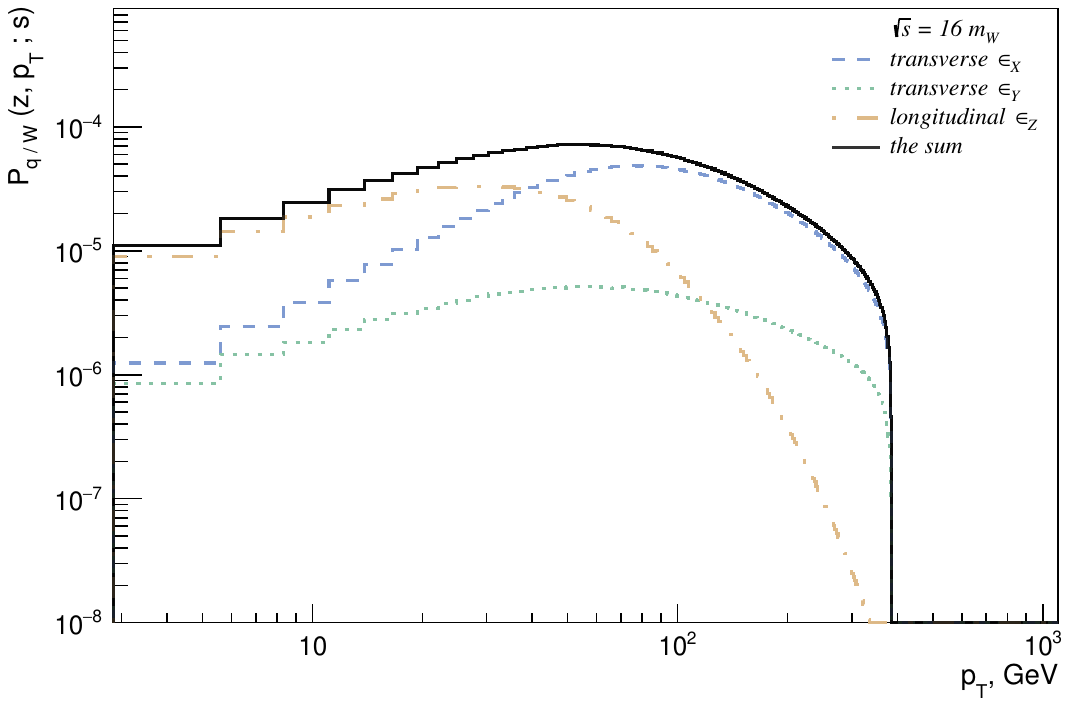}
\includegraphics[width=0.49\textwidth]{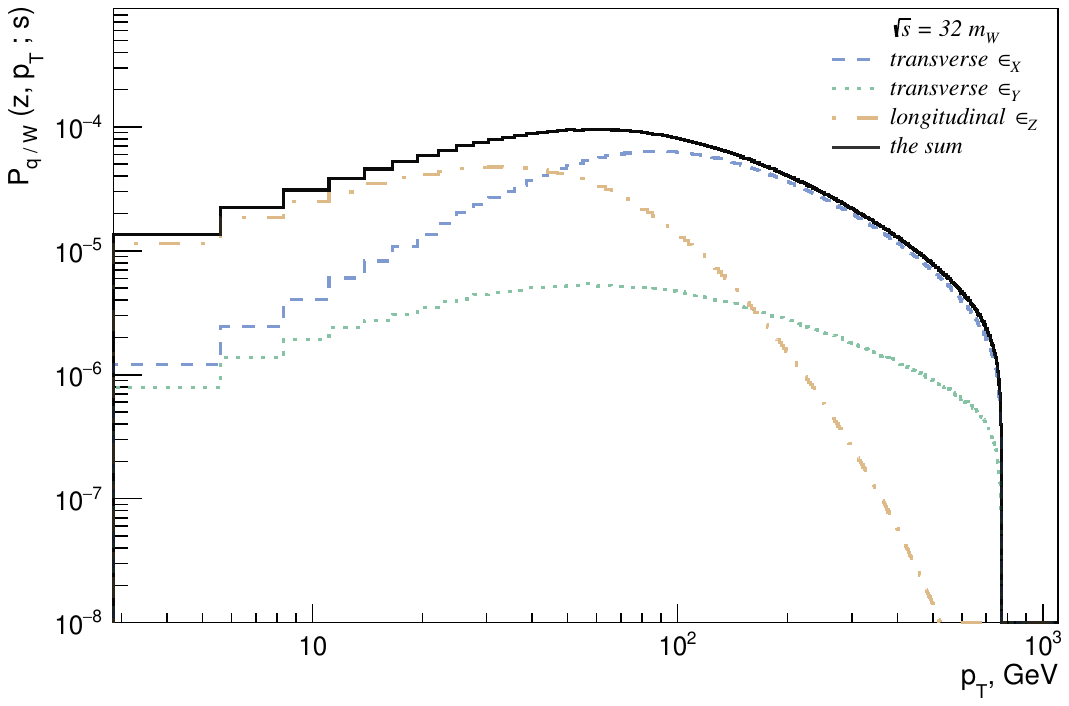}
\caption{
Net transverse momentum distributions obtained by integrating the splitting function 
$\P_{q/W}\,(z,p_T;s)$ over the allowed $z$ range for four different values of $s$.
Dashed curves, transverse polarization $\epsilon_{x}$ in the $Wq$ production plane;
dotted curves, transverse polarization $\epsilon_{y}$ perpendicular to the $Wq$ 
production plane; dash-dotted curves, longitudinal polarization $\epsilon_{z}$;
solid curves, the sum of all contributions. 
The values of $s$ from top to bottom: $\sqrt{s}=~ 4\,m_W,\,8\,m_W,\,16\,m_W,\,32\,m_W$.}
\label{fig:pt}
\end{figure}

\begin{figure}
\centering
\includegraphics[width=0.49\textwidth]{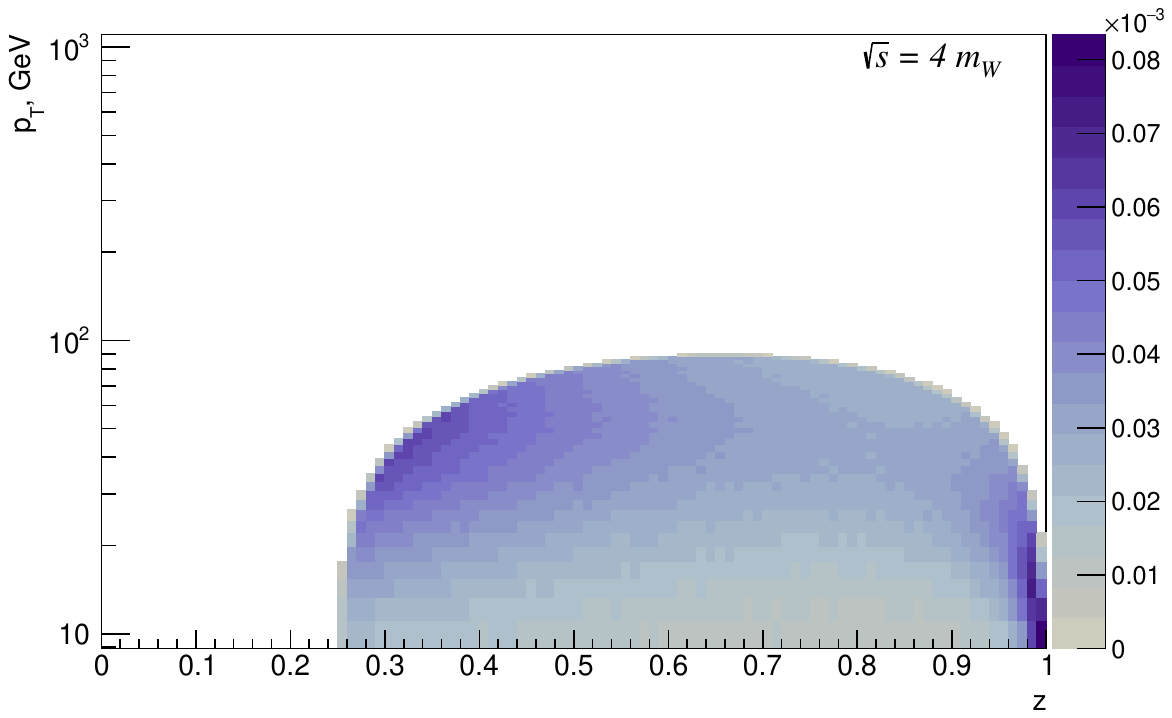}
\includegraphics[width=0.49\textwidth]{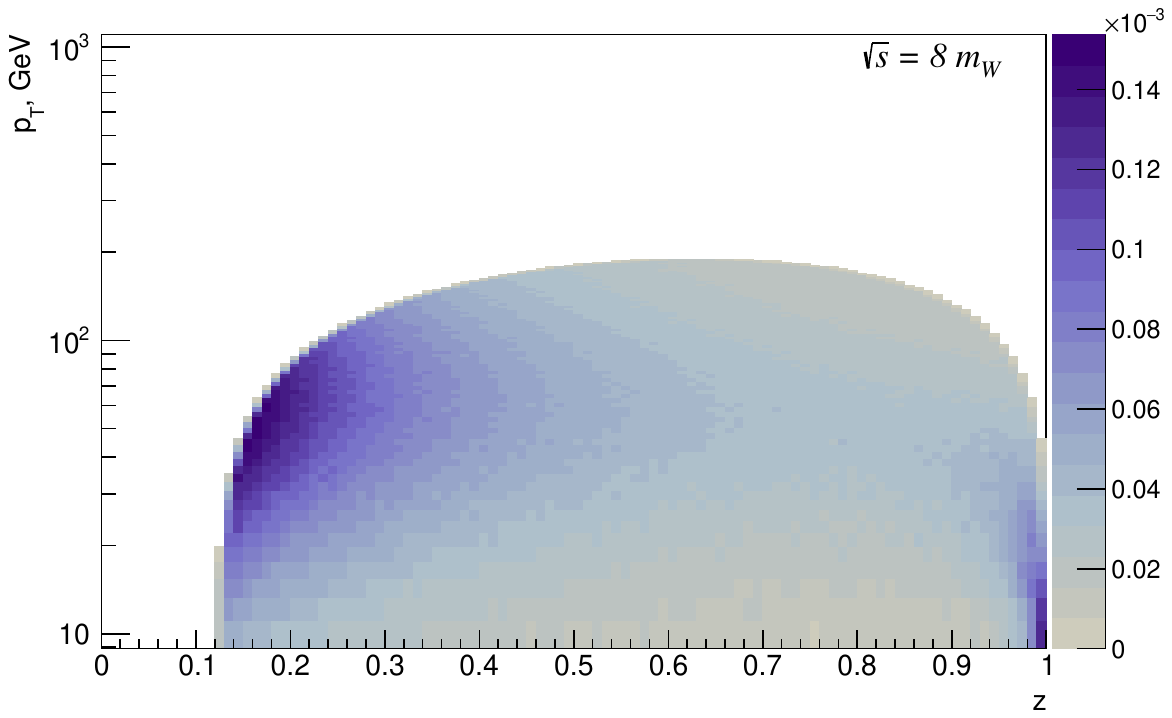}
\includegraphics[width=0.49\textwidth]{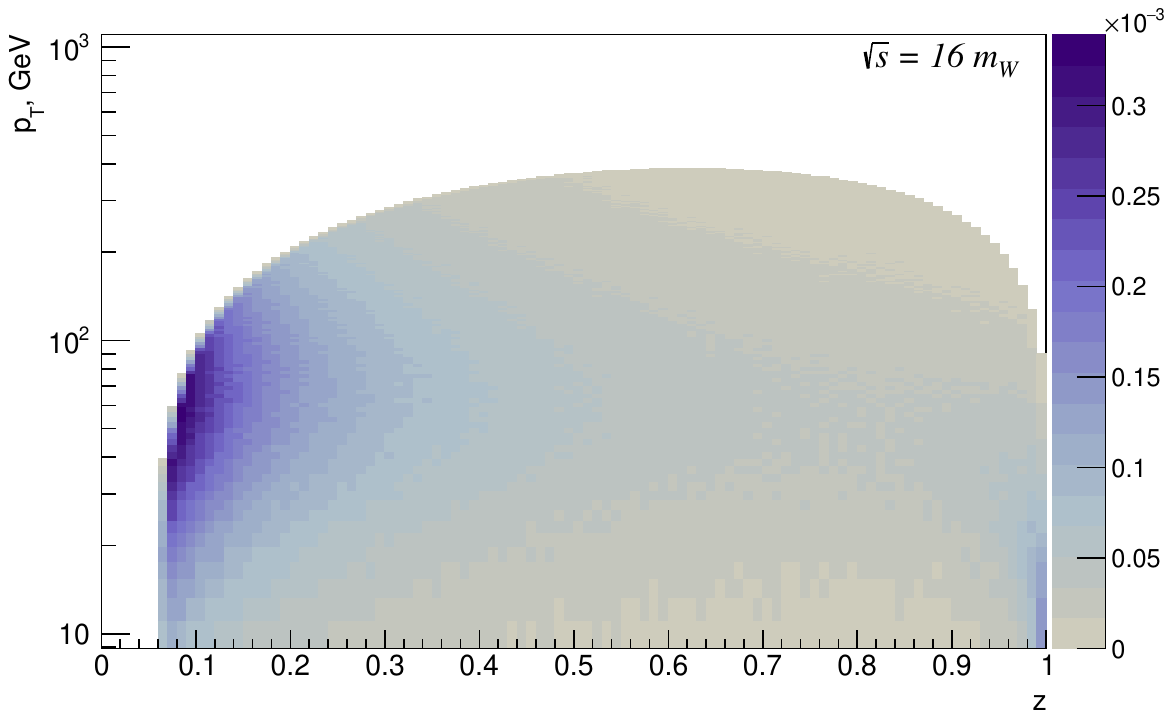}
\includegraphics[width=0.49\textwidth]{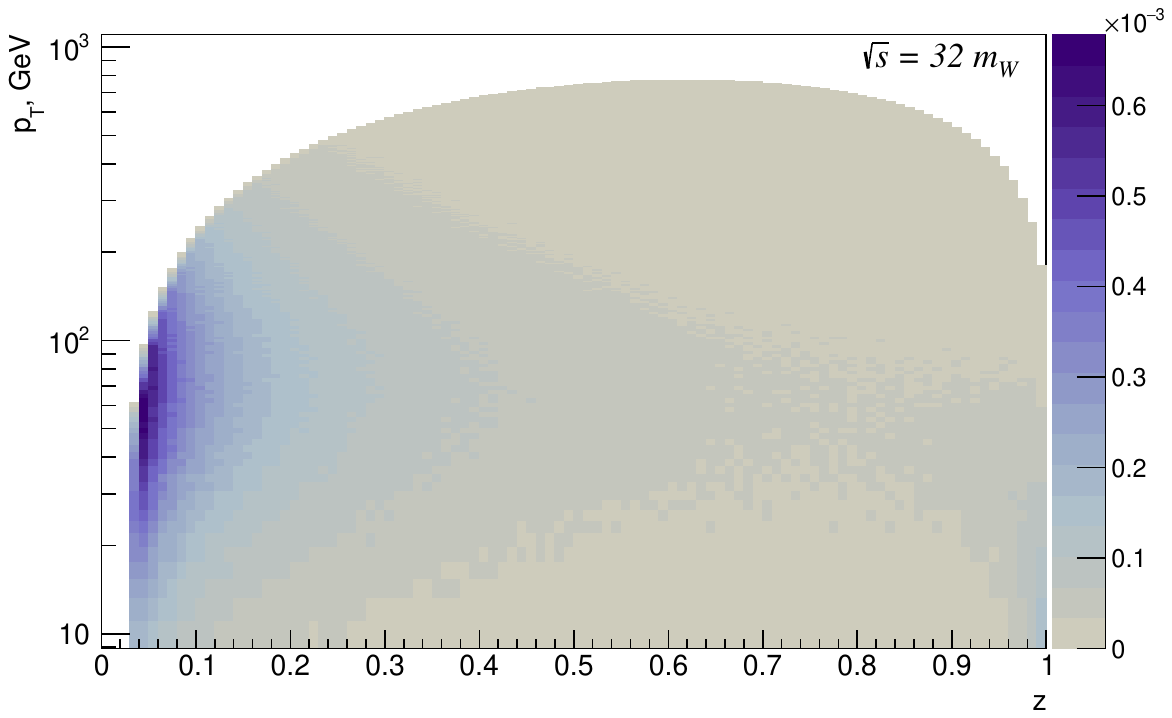}
\caption{Double differential distributions for $\P_{q/W}(z,p_T;s)$ summed over all polarizations.
The values of $s$ from top to bottom: $\sqrt{s}=~ 4\,m_W,\,8\,m_W,\,16\,m_W,\,32\,m_W$.}
\label{fig:ptz}
\end{figure}

Fig.~\ref{fig:fz} exhibits the behavior of $\P_{q/W}(z,p_T;s)$ as a function of $z$ 
(integrated over the allowed $p_T$ range), Fig.~\ref{fig:pt} exhibits the behavior of 
$\P_{q/W}(z,p_T;s)$ as a function of $p_T$ (integrated over the allowed $z$ range), 
and Fig.~\ref{fig:ptz} presents the double differential distributions.

In Fig.~\ref{fig:fz} and \ref{fig:pt} we separately show the individual contributions 
coming from three independent polarization states of $W$ boson. The longitudinal and 
transverse momentum distributions show strong dependence on the input energy $s$, which
roughly determines the energy of the radiating quark: $E^*\simeq |p^*|\simeq s/2$. 
At low $s$, the lack of phase space pushes the splitting function towards large $z$ and
makes the overall splitting probability small. At larger $s$, we see an evolution of the
splitting function towards smaller $z$ and an increase in its overall normalization. 
Finally, at very high $s$, it restores the shape of Weizs\"{a}cker-Williams approximation
$\P\,(z)\propto \bigl(1+(1-z)^2\bigr)/z$, in full consistency with the results of
Refs. \cite{Kane,Dawson,Sjostrand}.

\section{Summary}
\label{summ}
We have revisited the topic of effective $W^\pm, \,Z^0$ approximation paying attention
to three new aspects.
First, we considered the effective approximation as a function of two kinematic 
variables, $z$ and $p_T$. 
Second, we took into account kinematic restrictions connected with nonzero $W$ and $Z$
masses and nonzero $p_T$.
Third, we separately considered three different polarization states of the bosons.
We gained an interesting outcome from all these three points.\\
-- We have found that the transverse momentum is by far not negligible and can cause
a substantial deviation of the produced boson from the direction of the parent quark.\\
-- We have found that the phase space restrictions do dramatically affect the shape
and the overall normalization of the effective $W$ and $Z$ spectra.\\
-- We have found a significant difference between two transverse polarization
states of the produced bosons that can immediately manifest in the
angular distributions of the W and Z decay products.

So, we believe that our calculations provide a more accurate description 
of the event topology and particle spectra than it can be achieved with the ordinary 
WWA~splitting function.
Our expressions for $\P\,(z,p_T;s)$ are turned into a {\sc fortran} and a {\sc C++}
public codes which can be conveniently incorporated in Monte Carlo event generators.

\section*{Acknowledgements}
The authors thank Hannes Jung for stimulating their work.

\end{document}